\documentclass{article}
\usepackage{spconf,amsmath,graphicx}
\usepackage{amssymb} 
\usepackage{multirow}
\usepackage{rotating}
\usepackage{algorithmic} 
\usepackage{pseudocode} 
\usepackage{tipa} 
\usepackage{booktabs}
\usepackage{url}
\title{Parametric Representation for Singing Voice Synthesis: a Comparative Evaluation}
\name{Onur Babacan$^1$, Thomas Drugman$^1$, Tuomo Raitio$^2$, Daniel Erro$^3$, Thierry Dutoit$^1$
\thanks{O. Babacan is supported by a PhD grant funded by UMONS and Acapela Group. T. Drugman is supported by FNRS. T. Raitio is supported by the European Community's Seventh Framework Programme (FP7/2007-2013) under grant agreement n$^{\circ}$ 287678. The authors would like to thank Nathalie Henrich for the permission to use the LYRICS database.}}
\address{$^1$TCTS Lab - University of Mons, Belgium \\
$^2$Aalto University, Department of Signal Processing and Acoustics, Espoo, Finland\\
$^3$Ikerbasque - University of the Basque Country, Bilbao, Spain}

\begin{document}
\ninept
\maketitle
\begin{abstract}
Various parametric representations have been proposed to model the speech signal. While the performance of such vocoders is well-known in the context of speech processing, their extrapolation to singing voice synthesis might not be straightforward. The goal of this paper is twofold. First, a comparative subjective evaluation is performed across four existing techniques suitable for statistical parametric synthesis: traditional pulse vocoder, Deterministic plus Stochastic Model, Harmonic plus Noise Model and GlottHMM. The behavior of these techniques as a function of the singer type (baritone, counter-tenor and soprano) is studied. Secondly, the artifacts occurring in high-pitched voices are discussed and possible approaches to overcome them are suggested.
\end{abstract}
\begin{keywords}
Singing Voice, Parametric Representation, Vocoder, Synthesis.
\end{keywords}

\section{Introduction}\label{sec:Intro}
The field of singing synthesis has been steadily growing in maturity as many diverse techniques are being proposed and developed. Thanks to the similarities between singing and speech signals, techniques developed for speech synthesis undeniably influence the field, although direct applications have varying amounts of success due to some key differences. 
Some of the more significant differences of singing from speech are, the much wider pitch ranges, greater dynamic range, and significantly longer sustained voiced sounds.  The effects of source-filter interaction are greater and harder to neglect, in contrast to the commonly-made assumption in speech \cite{Titze2008}. Additionally, the diversity of singer categories and singing techniques makes it difficult to approach the problem of modeling singing in a straightforward manner, even when constrained to a single discipline of singing. As a direct consequence of such difficulties, many existing singing synthesizers have limited scope, generally working better for one singing technique or singer category at the expense of others. Therefore there exists a wide gap between the capabilities of existing synthesizers and the expressive range of human singers, as well as the performative requirements of musicians wishing to use these tools.

%paraphrase this
Among existing systems, Harmonic plus Noise Modeling (HNM) has been used extensively \cite{Stylianou05}. In the Vocaloid \cite{vocaloid} system, a degree of control is obtained over a unit concatenation technique by integrating HNM \cite{Bonada01}, though the synthesis results are still confined in singing space to the range of the pre-recorded samples. In the CHANT \cite{Rodet84} and FOF \cite{FOF} systems, rule-based descriptions characterizing some opera voices are integrated, yielding remarkable results for soprano voices. Meron obtained convincing results for lower registers of singing by applying the non-uniform unit selection technique to singing synthesis \cite{Meron00}. Similar strategies have been applied to formant synthesis, articulatory synthesis \cite{Birkholz06} and Hidden Markov Model (HMM)-based synthesis methods \cite{Saino06}, but the limitations in vocal expression range have been quite similarly limited.

Among the mentioned approaches, the HMM-based statistical parametric synthesis is of particular interest due to its flexibility and capability to be adapted to different circumstances. An immediate and fundamental question in this approach is the choice of parametric representation of signals. The limitations of any representation are unavoidably present in the synthesis, and often create the quality bottleneck. Many existing vocoders have been used to generate and synthesize from parameters modeled and generated by HMM-based signals. We can generally group the state-of-the-art vocoders into three categories with representative examples: {\it i) Source-filter with residual modeling:} Pulse vocoder, Deterministic plus Stochastic Model (DSM) \cite{DSM_TASLP}\cite{DSM_IS}, Closed-Loop Training \cite{Maia}, Mixed Excitation \cite{Yoshimura}, STRAIGHT \cite{Kawahara01}
{\it ii) Sinusoids+noise models:} Harmonic plus Noise Model (HNM) \cite{ErroSNH2013}, Harmonic/Stochastic Model (HSM) \cite{BanosEBM2008}, Sinusoidal Parametrization \cite{ShechtmanS2010}
{\it iii) Glottal modeling:} \mbox{GlottHMM} \cite{Raitio08} and variants \cite{Raitio_ICASSP11}\cite{Raitio}, Glottal Post-filtering \cite{cabralThesis}, Glottal Spectral Separation \cite{Cabral07}, Separation of Vocal-tract and Liljencrants-Fant model plus Noise (SVLN) \cite{LanchantinDR2010}.

As mentioned earlier, some of these vocoders have already been used in singing synthesizers. However, not all of them are suitable for statistical modeling, and their performance on singing is largely unknown. The goal of this paper is to evaluate the performance of a subset of these vocoders that are suitable for statistical modeling on a large variety of singing sounds by subjective listening tests, along with the conventional pulse vocoder to provide a baseline. %pesq? 
More specifically, DSM \cite{DSM_IS}, \mbox{GlottHMM} \cite{Raitio} and HNM \cite{ErroSNH2013} methods were selected for comparison. The choice of methods was motivated by covering different vocoder families.

The structure of the paper is as follows. Section 2 describes the vocoders selected for comparison. Section 3 presents the database used in the evaluation and the experimental protocol, as well as the results and their discussion. Section 4 concludes the paper.

%%%%%%%%%%%%%%%%
\section{Techniques for Singing Voice Parameterization}\label{sec:Techniques}

\subsection{Conventional Pulse Vocoder}
\label{ssec:Pulse}
This method is the simplest conventional framework used for parametric speech synthesis. It relies on a source-filter approach in which the excitation is either a Dirac pulse train when the signal is voiced, or a white noise for non-periodic segments. The filter is modeled in this study with Mel-Generalized Cepstral (MGC, \cite{MGC}) coefficients of order 24 with $\alpha = 0.42$ ($F_s=16$kHz) and $\gamma = 0$. Finally, the excitation is filtered with the mel-generalised log spectral approximation (MGLSA) filter \cite{Kobayashi85}.

\subsection{Deterministic plus Stochastic Model}
\label{ssec:DSM}
The Deterministic plus Stochastic Model (DSM) was proposed in \cite{DSM_IS,DSM_TASLP} to model the residual signal (obtained by inverse filtering after removing the contribution of the spectral envelope). DSM consists of two components acting in two distinct spectral sub-bands demarcated by the so-called \emph{maximum voiced frequency} (usually noted $F_m$): the deterministic contribution holds below $F_m$, while the stochastic component holds beyond $F_m$. These two contributions are fixed for a given speaker and are estimated by an analysis led on a speaker-dependent database. The deterministic component is defined as the first eigenvector obtained by Principal Component Analysis (PCA, \cite{PCA}) on a dataset of pitch-synchronous residual frames. Pitch marks were defined as the Glottal Closure Instants (GCIs) determined by the SEDREAMS algorithm \cite{SEDREAMS}. The resulting first eigenvector is then resampled to the target $F0$ value at synthesis time. The stochastic component is a white Gaussian noise further filtered to keep its content above $F_m$ and whose time structure is modulated by an Hilbert envelope estimated by averaging the noisy part of the same GCI-synchronous residual frames. Both deterministic and stochastic components are finally added and the resulting excitation signal filtered by the MGLSA filter with the same MGC coefficients as described in Section \ref{ssec:Pulse}.

In \cite{DSM_TASLP}, the maximum voiced frequency $F_m$ was fixed to a constant value. For neutral speech, this value turned out to be around 4 kHz. In singing voice, however, harmonics reach much higher frequencies and $F_m$ is fixed to 7 kHz for this study. This value comes from an inspection of various singing voice spectra. As a consequence, the input features of the DSM vocoder are the MGC coefficients for the filter and pitch ($F0$); all other data (like $F_m$, the first eigenvector or the noise envelope) being pre-estimated on the dataset of GCI-synchronous residual frames.

\subsection{Harmonic plus Noise Model}
\label{ssec:HNM}

This vocoder was extensively described in \cite{ErroSNH2013}. It parameterizes speech signals into three different streams: $F0$, a Mel-cepstral representation of the spectral envelope, and the maximum voiced frequency $F_m$. The vocoder includes an autocorrelation-based $F0$ estimation method. After refining the initial $F0$ estimate to meet the requirements of the subsequent algorithms, signals are analyzed by means of a full-band harmonic model. Then, the so called regularized discrete cepstrum technique \cite{CappeM1996} is applied to jointly interpolate between harmonic log-amplitudes and parameterize the resulting spectral envelope. The maximum voiced frequency estimation algorithm is based on a two-band partition of the analysis band according to the sinusoidal likeness of the spectral peaks therein. A smooth evolution of $F_m$ over time is imposed by means of a dynamic programming procedure.

Speech signals are reconstructed by overlapping short stationary frames consisting of a harmonic component and a noisy component. The amplitudes and phases that define the harmonic component are obtained by resampling the log-amplitude envelope and the minimum-phase envelopes given by the Mel-cepstral coefficients at multiples of $F0$ in the band [$0-F_m$]. An $F0$-dependent linear-in-frequency phase term is considered to guarantee the waveform coherence between adjacent frames. The noisy component is also built from the spectral envelope given by the Mel-cepstral coefficients. It is generated through inverse FFT after being modified in frequency by a piecewise linear high-pass filter with $F_m$ cut-off frequency. The noisy samples are finally time-modulated by means of a deterministic window.

In the experiments (see Section ~\ref{sec:Protocol}), we used the default configuration of the vocoder except for the $F0$ contour, which in this case was calculated from the EGG signal and supplied as an external input. The analysis period was 5~ms (the usual one in statistical parametric speech synthesis) and the order of the Mel-cepstral parameterization was 39. 
%No special care was taken regarding the particular nature of the signals under analysis.

\subsection{GlottHMM}
\label{ssec:GlottHMM}

GlottHMM \cite{Raitio,Raitio_ICASSP11} is a vocoder that uses glottal inverse filtering (GIF) in order to separate the speech signal into the vocal tract filter contribution and the voice source signal. Iterative adaptive inverse filtering (IAIF) \cite{Alku92} is used for GIF, inside which linear prediction (LP) is used for the estimation of the spectrum. IAIF is based on estimating and canceling the vocal tract filter and voice source spectral contributions using high and low order LP, respectively. The IAIF method produces an estimate of the voice source signal that is first used for estimating the fundamental frequency (F0) using autocorrelation method. Then, harmonic-to-noise ratio (HNR) of five frequency bands is estimated from the voice source signal by comparing the upper and lower smoothed spectral envelopes constructed from the harmonic peaks and the interharmonic valleys, respectively. In the case of voiced speech, GCIs are detected from the differentiated glottal flow signal using simple peak picking of prominent negative values in the signal at fundamental period intervals. GCIs are then used for pitch-synchronous analysis of the speech signal, where IAIF is applied again for each (overlapping) two-pitch period speech segments to produce new estimates for the vocal tract spectrum and the voice source segment. The pitch-synchronous analysis is performed in order to reduce the interfering effect of the excitation harmonics to the vocal tract spectrum, which is especially important in high-pitched singing voice. From each pitch-synchronous segment, a vocal tract estimate is obtained, and the one being closest to the mean of all estimates in a frame is selected as the final vocal tract estimate. Similarly, the spectral contribution of each pitch-synchronous segment is estimated using low-order LP, and the final estimate is the closest to the mean in a frame. Both of these spectral features are further converted to line spectral frequencies (LSF) \cite{Soong84} in order to achieve a better parameter representation for a possible subsequent statistical modeling. The energy of the speech signal is evaluated from the original speech frame. 

In synthesis, a pre-stored natural glottal flow pulse is used for creating the voiced excitation. First, the pulse is interpolated to achieve a desired duration according to $F_0$ and scaled in energy according to the energy measure. In order to control the degree of voicing, the excitation signal is mixed with noise in each frequency band according to the band-wise HNR. In order to control the phonation characteristics, the spectrum of the excitation is matched with the voice source LP spectrum. Finally, the excitation is fed to the vocal tract filter to create speech.

%%%%%%%%%%%%%%%%
\section{Experiments}\label{sec:Protocol}

\subsection{Data}
%paraphrase
For this study, the scope was constrained to vowels. Samples with verified reference pitch trajectories from our previous study were used \cite{babacanPitch}. Samples of  different singers were taken from the LYRICS database \cite{Henrich2001,Henrich2005}, for a  total of 13 trained singers. The selection consisted of 7 bass-baritones, 3 countertenors, and 3 sopranos. The recording sessions took place in a soundproof booth. Acoustic and electroglottographic signals were recorded simultaneously on the two channels of a DAT recorder. The acoustic signal was recorded using a condenser microphone (Br\"{u}el \& Kj\ae r 4165) placed 50 cm from the singer's mouth, a preamplifier (Br\"{u}el \& Kj\ae r 2669), and a conditioning amplifier (Br\"{u}el \& Kj\ae r NEXUS 2690). The electroglottographic signal was recorded using a two-channel electroglottograph (EG2, \cite{Rothenberg1992}). The selected samples contain a variety of singing tasks, such as sustained vowels, crescendos-decrescendos and arpeggios, and ascending and descending glissandos. %Whenever possible, the singers were asked to sing in both laryngeal mechanisms M1 and M2 \cite{Henrich2006,Roubeau2009}. Laryngeal mechanisms M1 and M2 are two biomechanical configurations of the laryngeal vibrator commonly used in speech and singing by both male and females. Basses and baritones mainly sing in M1. They may use M2 to sing high pitches. Countertenors commonly sing in M2. They may use M1 for artistic purposes, so their singing technique requires the ability to sing with similar voice qualities in both laryngeal mechanisms. Sopranos mainly sing in M2. They can choose to sing in M1 in the low to medium part of their tessitura.

\subsection{Subjective Evaluation}

A Comparison Mean Opinion Score (CMOS) test was conducted online for the four vocoders described in Section ~\ref{sec:Techniques} with the parameters as given in Table ~\ref{tab:param}. Where necessary, pitch values were supplied from the ground truth established in \cite{babacanPitch} in order to eliminate any discrepancies between vocoders due to different pitch tracking results.

Sixteen participants of expert and non-expert backgrounds took part in the test. Given a reference sample and two copy-synthesis samples A and B from different vocoders, the participants were asked to compare the two and decide whether "A is much better/better/slightly better/about the same/slightly worse/worse/much worse than B". The scale is represented by integers in [-3 3] in the scoring. Three singer types in the database were evaluated: baritone, counter-tenor and soprano. The four methods compared create a total of six unique pairings, and two questions were asked per pairing and per singer category, for a total of 36 questions per participant, resulting in 96 scores per singer category for each method. The samples used in questions were selected and arranged randomly for each participant.

\begin{table}[b!] 
\caption{Acoustic features used by the various vocoders.} 
\begin{center} 
\begin{tabular}{c|c|c} 
\hline 
Vocoder & Feature & \# of param. \\ 
\hline 
Pulse & F0 & 1 \\ 
 & MGC coefficients & 25 \\ 
\hline 
DSM & F0 & 1 \\ 
 & MGC coefficients & 25 \\ 
\hline 
HNM & F0 & 1 \\ 
 & Mel Cepstral coefficients & 40 \\
 & Maximum Voice Frequency (Fm) & 1 \\ 
\hline 
GlottHMM & Energy & 1 \\ 
 & F0 & 1 \\ 
 & HNRs & 5 \\ 
 & Voice source spectrum & 10 \\ 
 & Vocal tract spectrum & 30 \\ 
\hline 
\end{tabular} 
\label{tab:param} 
\end{center} 
\end{table}

\subsection{Results}\label{ssec:Results}

\begin{figure*}
  \vspace{-12pt}
  \centering
  \includegraphics[width=\linewidth]{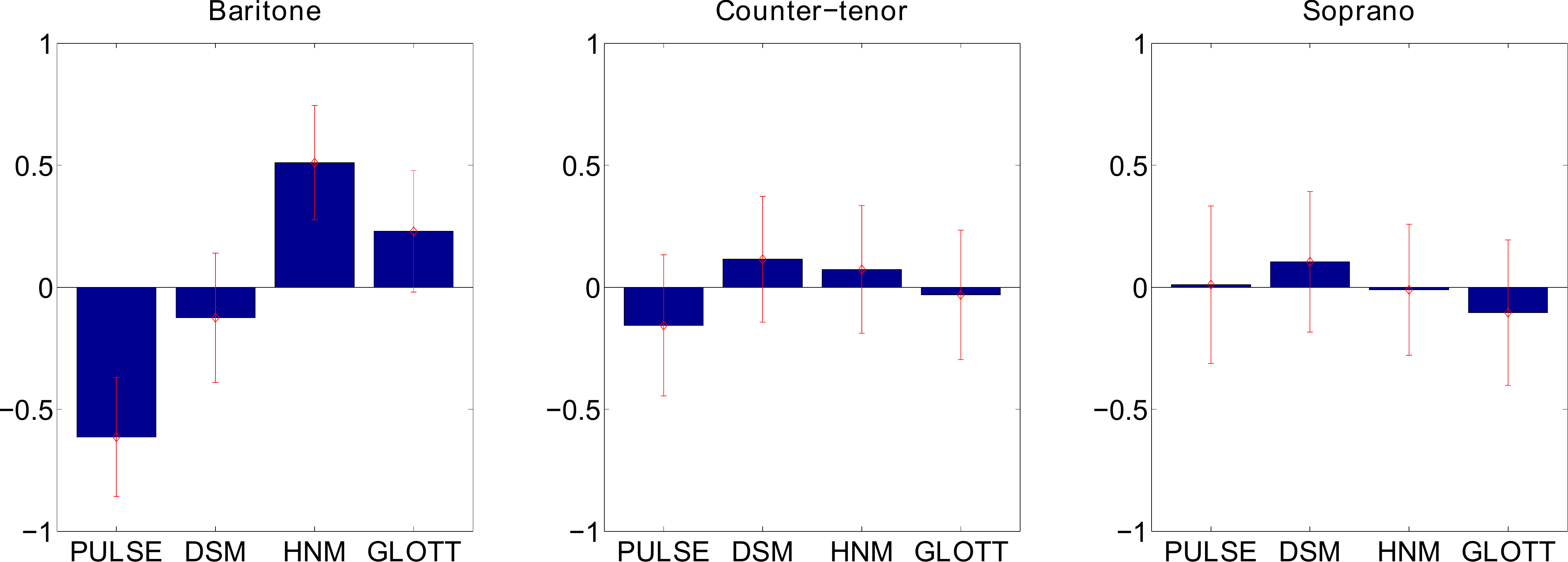}	
  \caption{Average CMOS scores for the 4 compared techniques and per singer category, together with their 95\% confidence intervals.}
  \label{fig:Results}  
	\vspace{-3pt}  
\end{figure*}

The average results per singer category are displayed in Figure \ref{fig:Results}. It can be observed that the relative performance of the techniques is highly dependent upon the pitch. While differences across methods are notable for baritones, they turn out to become much more reduced for high-pitched voices. For baritones, the conventional pulse vocoder clearly gives the worst quality and is outperformed by the three other techniques. The best vocoder appears to be HNM, followed respectively by GlottHMM and DSM. The results obtained for this group seem to be consistent with the evaluation done in \cite{Hu2013}, where the order of preference for a male voice was the same. For singers using higher $F0$ values, the ranking across techniques gets dramatically altered, although differences between the techniques are no longer statistically significant. Among others, it is worth noting that the improvement over the traditional pulse vocoder becomes marginal. This reduction can be explained in several ways. First, the increase of $F0$ in high-pitched voices generally goes along with an increase of the maximum voiced frequency $F_m$. As a consequence, the noise modeling becomes less and less important, which explains the fact that the difference between Pulse and DSM becomes less substantial. Secondly, the performance of IAIF (or any other method) in estimating the glottal flow is known to get degraded as $F0$ increases \cite{DrugmanGFComparison}. This partly justifies the drop of quality for GlottHMM. Finally, as further analyzed in Section \ref{ssec:Discussion}, $F_m$ estimation becomes problematic in high-pitched singing voices. As a result, the quality of HNM (which is based on a dynamic $F_m$) is affected, contrarily to DSM which makes use of a static $F_m$ (fixed to 7 kHz).

\subsection{Discussion and perspectives}\label{ssec:Discussion}
A careful listening and inspection of the signals revealed that high-pitched singing voices are more prone to artifacts. More precisely, we could identify three main possible sources of degradation for such voices, illustrated with the help of Figure \ref{fig:artifact_Illustration}:

\begin{itemize}
\item Some harmonics below the actual $F_m$ are almost inexistent which could lead to an underestimation of the maximum voiced frequency. Considering the example of Fig. \ref{fig:artifact_Illustration} (see the top panel), the $4^{th}$ and $5^{th}$ harmonics have a very low amplitude while spectral peaks in the $6^{th}$ and $7^{th}$ harmonics (and even further) clearly emerge. This is likely to result in an underestimated $F_m$ using the algorithms described in \cite{StylianouThesis} or \cite{ErroSNH2013}. As a consequence, any vocoder based on a dynamic maximum voiced frequency will sound artificially too noisy in such segments where $F_m$ is underestimated.

\item The MGC spectral envelope captures $F0$-related information for high-pitched voices. This can be observed in the dashed line in the top plot of Fig. \ref{fig:artifact_Illustration}, which exhibits clear resonances for the first three harmonics. Although this effect is not critical in a copy-synthesis scheme (as achieved in this paper), this will undoubtedly become problematic when applying pitch transposition, or when having HMM-based speech synthesis in view. Indeed, the passage of a periodic excitation signal at a different $F0$ value through such a filter will cause auditory artifacts: the synthesized signal will contain a double pitch, since the filter contains residual pitch information. 

\item The spectrum may exhibit relatively strong interharmonics. This can be for example noticed by spectral peaks in $7/2 \cdot F0$ and $13/2 \cdot F0$ in the top plot of Fig. \ref{fig:artifact_Illustration}. These peaks have an amplitude comparable to their neighbouring harmonics. The implication of these interharmonics might be threefold: \emph{i)} vocoders considered in this work are only able to reproduce harmonics and will fail in modeling such interharmonics; \emph{ii)} it might affect the $F_m$ estimation process; \emph{iii)} although not taken into account in this study (as we consider the $F0$ ground truth from EGG recordings), this will have an impact on the performance of pitch tracking methods, such as the Summation of Residual Harmonics (SRH, \cite{SRH}) algorithm. The spectrum of the residual signal is displayed for information in the bottom plot of Fig \ref{fig:artifact_Illustration}. Strong peaks in interharmonics can be observed; these are notably due to the inappropriate spectral envelope.  Note that the physiological origin of these interharmonic peaks would require further investigation.
\end{itemize}

These issues should be alleviated in order to have a high-quality parametric representation of singing voice. First, the spectral weighting of the noisy component should either be based on appropriate aperiodicity measurements \cite{ZenTNT2007}, HNR \cite{Raitio} or bandpass voicing strengths \cite{Yoshimura} in different spectral subbands, or involve the use of a dynamic $F_m$ whose values are estimated by a new algorithm specifically designed to overcome the first and third aforementioned problems. Secondly, conventional MGC extraction turns out to be inappropriate unless the cepstral order is dynamically adjusted according to $F0$, which is not practical when signals exhibit large $F0$ variations. Even considering spectral analysis techniques where the effect of harmonicity is eliminated before cepstral fitting \cite{Roebel,ZenTNT2007,CappeM1996,ErroSNH2013}, the assumption of a harmonic or quasi-harmonic spectral structure may be insufficient for two main factors: (i) the low spectral resolution when $F0$ is high, which could result in poor spectral envelope estimation and inconsistent measurements between adjacent frames; (ii) the presence of the aforementioned interharmonic tones. The latter problem is ignored by current speech parameterization systems. However, its perceptual importance should be assessed in the context of singing voices. If this is revealed to be crucial, a vocoder offering the possibility to generate interharmonics and to deal with them during analysis should be developed.
%Therefore, MGC should be replaced by techniques such as the True Envelope \cite{Imai}, \cite{Roebel} or possibly applying a low-pass cepstral liftering below the fundamental quefrency. Finally, the perceptual importance of interharmonics should be assessed. If this reveals to be crucial, a vocoder offering the possibility to generate interharmonics should be developed.  

section{Parametric Representation of Singing Voice}

\subsection{Introduction}

\subsection{Techniques for Parametric Representation of Singing Voice}

\subsubsection{Conventional Pulse Vocoder}
\label{sssec:Pulse}
This method is the simplest conventional framework used for parametric speech synthesis. It relies on a source-filter approach in which the excitation is either a Dirac pulse train when the signal is voiced, or a white noise for non-periodic segments. The filter is modeled in this study with Mel-Generalized Cepstral (MGC, \cite{MGC}) coefficients of order 24 with $\alpha = 0.42$ ($F_s=16$kHz) and $\gamma = 0$. Finally, the excitation is filtered with the mel-generalised log spectral approximation (MGLSA) filter \cite{Kobayashi85}.

\subsubsection{Deterministic plus Stochastic Model}
\label{sssec:DSM}
The Deterministic plus Stochastic Model (DSM) was proposed in \cite{DSM_IS,DSM_TASLP} to model the residual signal (obtained by inverse filtering after removing the contribution of the spectral envelope). DSM consists of two components acting in two distinct spectral sub-bands demarcated by the so-called \emph{maximum voiced frequency} (usually noted $F_m$): the deterministic contribution holds below $F_m$, while the stochastic component holds beyond $F_m$. These two contributions are fixed for a given speaker and are estimated by an analysis led on a speaker-dependent database. The deterministic component is defined as the first eigenvector obtained by Principal Component Analysis (PCA, \cite{PCA}) on a dataset of pitch-synchronous residual frames. Pitch marks were defined as the Glottal Closure Instants (GCIs) determined by the SEDREAMS algorithm \cite{SEDREAMS}. The resulting first eigenvector is then resampled to the target $F0$ value at synthesis time. The stochastic component is a white Gaussian noise further filtered to keep its content above $F_m$ and whose time structure is modulated by an Hilbert envelope estimated by averaging the noisy part of the same GCI-synchronous residual frames. Both deterministic and stochastic components are finally added and the resulting excitation signal filtered by the MGLSA filter with the same MGC coefficients as described in Section \ref{ssec:Pulse}.

In \cite{DSM_TASLP}, the maximum voiced frequency $F_m$ was fixed to a constant value. For neutral speech, this value turned out to be around 4 kHz. In singing voice, however, harmonics reach much higher frequencies and $F_m$ is fixed to 7 kHz for this study. This value comes from an inspection of various singing voice spectra. As a consequence, the input features of the DSM vocoder are the MGC coefficients for the filter and pitch ($F0$); all other data (like $F_m$, the first eigenvector or the noise envelope) being pre-estimated on the dataset of GCI-synchronous residual frames.

\subsubsection{Harmonic plus Noise Model}
\label{sssec:HNM}

This vocoder was extensively described in \cite{ErroSNH2013}. It parameterizes speech signals into three different streams: $F0$, a Mel-cepstral representation of the spectral envelope, and the maximum voiced frequency $F_m$. The vocoder includes an autocorrelation-based $F0$ estimation method. After refining the initial $F0$ estimate to meet the requirements of the subsequent algorithms, signals are analyzed by means of a full-band harmonic model. Then, the so called regularized discrete cepstrum technique \cite{CappeM1996} is applied to jointly interpolate between harmonic log-amplitudes and parameterize the resulting spectral envelope. The maximum voiced frequency estimation algorithm is based on a two-band partition of the analysis band according to the sinusoidal likeness of the spectral peaks therein. A smooth evolution of $F_m$ over time is imposed by means of a dynamic programming procedure.

Speech signals are reconstructed by overlapping short stationary frames consisting of a harmonic component and a noisy component. The amplitudes and phases that define the harmonic component are obtained by resampling the log-amplitude envelope and the minimum-phase envelopes given by the Mel-cepstral coefficients at multiples of $F0$ in the band [$0-F_m$]. An $F0$-dependent linear-in-frequency phase term is considered to guarantee the waveform coherence between adjacent frames. The noisy component is also built from the spectral envelope given by the Mel-cepstral coefficients. It is generated through inverse FFT after being modified in frequency by a piecewise linear high-pass filter with $F_m$ cut-off frequency. The noisy samples are finally time-modulated by means of a deterministic window.

In the experiments (see Section ~\ref{sec:Protocol}), we used the default configuration of the vocoder except for the $F0$ contour, which in this case was calculated from the EGG signal and supplied as an external input. The analysis period was 5~ms (the usual one in statistical parametric speech synthesis) and the order of the Mel-cepstral parameterization was 39. 
%No special care was taken regarding the particular nature of the signals under analysis.

\subsubsection{GlottHMM}
\label{sssec:GlottHMM}

GlottHMM \cite{Raitio,Raitio_ICASSP11} is a vocoder that uses glottal inverse filtering (GIF) in order to separate the speech signal into the vocal tract filter contribution and the voice source signal. Iterative adaptive inverse filtering (IAIF) \cite{Alku92} is used for GIF, inside which linear prediction (LP) is used for the estimation of the spectrum. IAIF is based on estimating and canceling the vocal tract filter and voice source spectral contributions using high and low order LP, respectively. The IAIF method produces an estimate of the voice source signal that is first used for estimating the fundamental frequency (F0) using autocorrelation method. Then, harmonic-to-noise ratio (HNR) of five frequency bands is estimated from the voice source signal by comparing the upper and lower smoothed spectral envelopes constructed from the harmonic peaks and the interharmonic valleys, respectively. In the case of voiced speech, GCIs are detected from the differentiated glottal flow signal using simple peak picking of prominent negative values in the signal at fundamental period intervals. GCIs are then used for pitch-synchronous analysis of the speech signal, where IAIF is applied again for each (overlapping) two-pitch period speech segments to produce new estimates for the vocal tract spectrum and the voice source segment. The pitch-synchronous analysis is performed in order to reduce the interfering effect of the excitation harmonics to the vocal tract spectrum, which is especially important in high-pitched singing voice. From each pitch-synchronous segment, a vocal tract estimate is obtained, and the one being closest to the mean of all estimates in a frame is selected as the final vocal tract estimate. Similarly, the spectral contribution of each pitch-synchronous segment is estimated using low-order LP, and the final estimate is the closest to the mean in a frame. Both of these spectral features are further converted to line spectral frequencies (LSF) \cite{Soong84} in order to achieve a better parameter representation for a possible subsequent statistical modeling. The energy of the speech signal is evaluated from the original speech frame. 

In synthesis, a pre-stored natural glottal flow pulse is used for creating the voiced excitation. First, the pulse is interpolated to achieve a desired duration according to $F_0$ and scaled in energy according to the energy measure. In order to control the degree of voicing, the excitation signal is mixed with noise in each frequency band according to the band-wise HNR. In order to control the phonation characteristics, the spectrum of the excitation is matched with the voice source LP spectrum. Finally, the excitation is fed to the vocal tract filter to create speech.

\subsubsection{STRAIGHT}

\subsection{Experimental Protocol}
\subsubsection{Subjective Evaluation}

A Comparison Mean Opinion Score (CMOS) test was conducted online for the four vocoders described in Section ~\ref{sec:Techniques} with the parameters as given in Table ~\ref{tab:param}. Where necessary, pitch values were supplied from the ground truth established in \cite{babacanPitch} in order to eliminate any discrepancies between vocoders due to different pitch tracking results.

Sixteen participants of expert and non-expert backgrounds took part in the test. Given a reference sample and two copy-synthesis samples A and B from different vocoders, the participants were asked to compare the two and decide whether "A is much better/better/slightly better/about the same/slightly worse/worse/much worse than B". The scale is represented by integers in [-3 3] in the scoring. Three singer types in the database were evaluated: baritone, counter-tenor and soprano. The four methods compared create a total of six unique pairings, and two questions were asked per pairing and per singer category, for a total of 36 questions per participant, resulting in 96 scores per singer category for each method. The samples used in questions were selected and arranged randomly for each participant.

\begin{table}[b!] 
\caption{Acoustic features used by the various vocoders.} 
\begin{center} 
\begin{tabular}{c|c|c} 
\hline 
Vocoder & Feature & \# of param. \\ 
\hline 
Pulse & F0 & 1 \\ 
 & MGC coefficients & 25 \\ 
\hline 
DSM & F0 & 1 \\ 
 & MGC coefficients & 25 \\ 
\hline 
HNM & F0 & 1 \\ 
 & Mel Cepstral coefficients & 40 \\
 & Maximum Voice Frequency (Fm) & 1 \\ 
\hline 
GlottHMM & Energy & 1 \\ 
 & F0 & 1 \\ 
 & HNRs & 5 \\ 
 & Voice source spectrum & 10 \\ 
 & Vocal tract spectrum & 30 \\ 
\hline 
\end{tabular} 
\label{tab:param} 
\end{center} 
\end{table}

\subsubsection{Results}\label{ssec:Results}

The average results per singer category are displayed in Figure \ref{fig:Results}. It can be observed that the relative performance of the techniques is highly dependent upon the pitch. While differences across methods are notable for baritones, they turn out to become much more reduced for high-pitched voices. For baritones, the conventional pulse vocoder clearly gives the worst quality and is outperformed by the three other techniques. The best vocoder appears to be HNM, followed respectively by GlottHMM and DSM. The results obtained for this group seem to be consistent with the evaluation done in \cite{Hu2013}, where the order of preference for a male voice was the same. For singers using higher $F0$ values, the ranking across techniques gets dramatically altered, although differences between the techniques are no longer statistically significant. Among others, it is worth noting that the improvement over the traditional pulse vocoder becomes marginal. This reduction can be explained in several ways. First, the increase of $F0$ in high-pitched voices generally goes along with an increase of the maximum voiced frequency $F_m$. As a consequence, the noise modeling becomes less and less important, which explains the fact that the difference between Pulse and DSM becomes less substantial. Secondly, the performance of IAIF (or any other method) in estimating the glottal flow is known to get degraded as $F0$ increases \cite{DrugmanGFComparison}. This partly justifies the drop of quality for GlottHMM. Finally, as further analyzed in Section \ref{ssec:Discussion}, $F_m$ estimation becomes problematic in high-pitched singing voices. As a result, the quality of HNM (which is based on a dynamic $F_m$) is affected, contrarily to DSM which makes use of a static $F_m$ (fixed to 7 kHz).

%\begin{figure*}
  %\vspace{-12pt}
  %\centering
  %\includegraphics[width=\linewidth]{Results.pdf}	
  %\caption{Average CMOS scores for the 4 compared techniques and per singer category, together with their 95\% confidence intervals.}
  %\label{fig:Results}  
	%\vspace{-3pt}  
%\end{figure*}

\subsubsection{Discussion and perspectives}\label{ssec:Discussion}
\begin{figure}
  %\vspace{-12pt}
  \centering
  \includegraphics[width=0.48\textwidth]{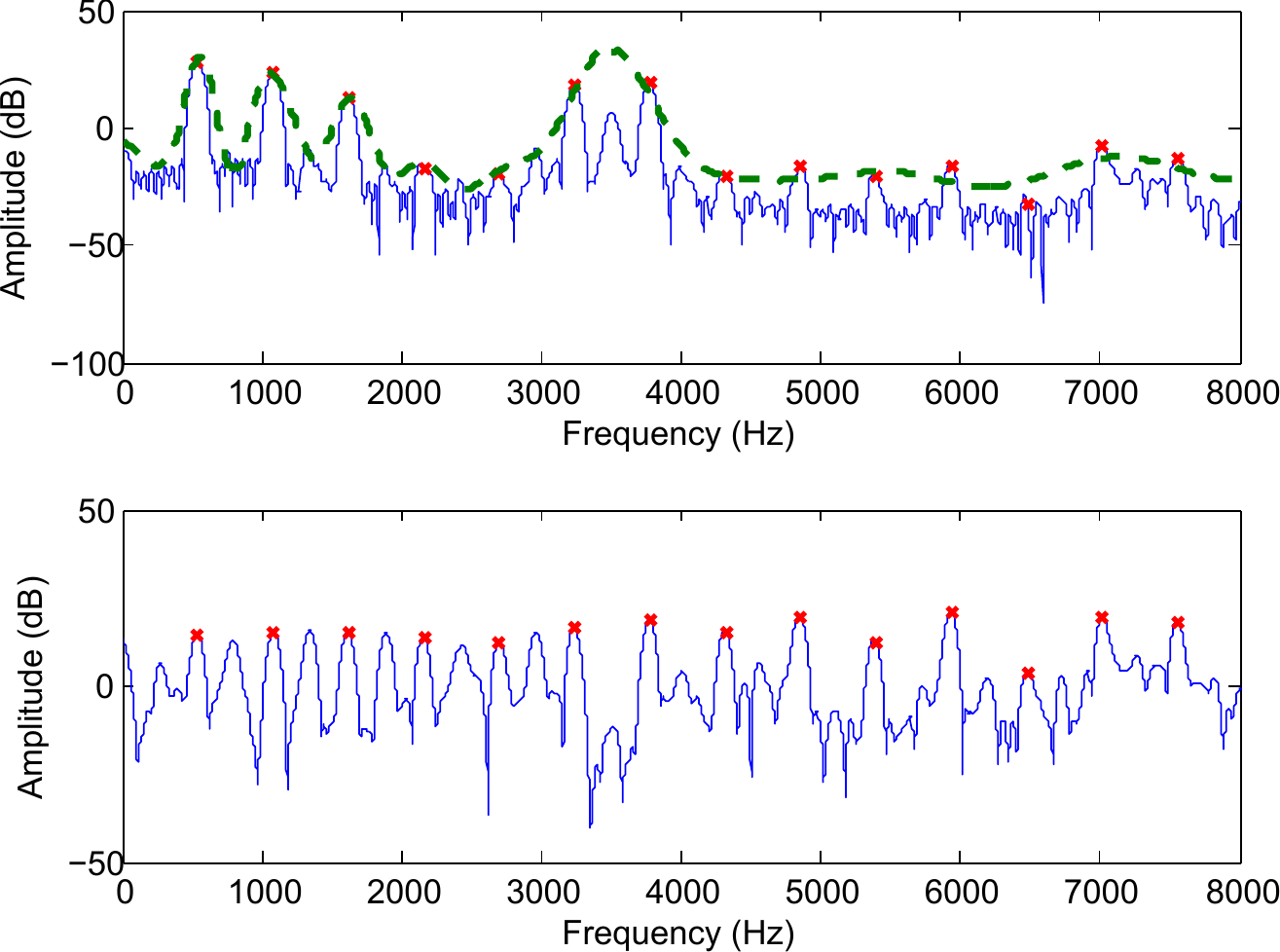}
	\vspace{-16pt}
  \caption{Example of analysis of a frame of singing voice from a soprano. \emph{Top panel:} its amplitude spectrum together with the harmonics (indicated by crosses) and the MGC spectral envelope (in dashed line); \emph{Bottom panel:} the spectrum of the corresponding residual signal obtained by MGC inverse filtering.}
  \label{fig:artifact_Illustration}  
	\vspace{-3pt}  
\end{figure}

A careful listening and inspection of the signals revealed that high-pitched singing voices are more prone to artifacts. More precisely, we could identify three main possible sources of degradation for such voices, illustrated with the help of Figure \ref{fig:artifact_Illustration}:

\begin{itemize}
\item Some harmonics below the actual $F_m$ are almost inexistent which could lead to an underestimation of the maximum voiced frequency. Considering the example of Fig. \ref{fig:artifact_Illustration} (see the top panel), the $4^{th}$ and $5^{th}$ harmonics have a very low amplitude while spectral peaks in the $6^{th}$ and $7^{th}$ harmonics (and even further) clearly emerge. This is likely to result in an underestimated $F_m$ using the algorithms described in \cite{StylianouThesis} or \cite{ErroSNH2013}. As a consequence, any vocoder based on a dynamic maximum voiced frequency will sound artificially too noisy in such segments where $F_m$ is underestimated.

\item The MGC spectral envelope captures $F0$-related information for high-pitched voices. This can be observed in the dashed line in the top plot of Fig. \ref{fig:artifact_Illustration}, which exhibits clear resonances for the first three harmonics. Although this effect is not critical in a copy-synthesis scheme (as achieved in this paper), this will undoubtedly become problematic when applying pitch transposition, or when having HMM-based speech synthesis in view. Indeed, the passage of a periodic excitation signal at a different $F0$ value through such a filter will cause auditory artifacts: the synthesized signal will contain a double pitch, since the filter contains residual pitch information. 

\item The spectrum may exhibit relatively strong interharmonics. This can be for example noticed by spectral peaks in $7/2 \cdot F0$ and $13/2 \cdot F0$ in the top plot of Fig. \ref{fig:artifact_Illustration}. These peaks have an amplitude comparable to their neighbouring harmonics. The implication of these interharmonics might be threefold: \emph{i)} vocoders considered in this work are only able to reproduce harmonics and will fail in modeling such interharmonics; \emph{ii)} it might affect the $F_m$ estimation process; \emph{iii)} although not taken into account in this study (as we consider the $F0$ ground truth from EGG recordings), this will have an impact on the performance of pitch tracking methods, such as the Summation of Residual Harmonics (SRH, \cite{SRH}) algorithm. The spectrum of the residual signal is displayed for information in the bottom plot of Fig \ref{fig:artifact_Illustration}. Strong peaks in interharmonics can be observed; these are notably due to the inappropriate spectral envelope.  Note that the physiological origin of these interharmonic peaks would require further investigation.
\end{itemize}

These issues should be alleviated in order to have a high-quality parametric representation of singing voice. First, the spectral weighting of the noisy component should either be based on appropriate aperiodicity measurements \cite{ZenTNT2007}, HNR \cite{Raitio} or bandpass voicing strengths \cite{Yoshimura} in different spectral subbands, or involve the use of a dynamic $F_m$ whose values are estimated by a new algorithm specifically designed to overcome the first and third aforementioned problems. Secondly, conventional MGC extraction turns out to be inappropriate unless the cepstral order is dynamically adjusted according to $F0$, which is not practical when signals exhibit large $F0$ variations. Even considering spectral analysis techniques where the effect of harmonicity is eliminated before cepstral fitting \cite{Roebel,ZenTNT2007,CappeM1996,ErroSNH2013}, the assumption of a harmonic or quasi-harmonic spectral structure may be insufficient for two main factors: (i) the low spectral resolution when $F0$ is high, which could result in poor spectral envelope estimation and inconsistent measurements between adjacent frames; (ii) the presence of the aforementioned interharmonic tones. The latter problem is ignored by current speech parameterization systems. However, its perceptual importance should be assessed in the context of singing voices. If this is revealed to be crucial, a vocoder offering the possibility to generate interharmonics and to deal with them during analysis should be developed.

%%%%%%%%%%%%%%%%%
%%%%%%%%%%%%%%%%%s_
\section{Conclusion}
\label{sec:Conclu}
In this paper, a subjective evaluation of three state-of-the-art vocoders and a baseline vocoder was made on a variety of singing sounds, as a function of singer type (baritone, counter-tenor and soprano). Listener preferences were presented. It was observed that increasing fundamental frequency creates different problems for all vocoders, and the preference between them becomes statistically insignificant due to these quality degradations. According to the results of the study, the current vocoders need improvements in the aperiodicity estimation of high-pitched voice and in spectral estimation free of the effect of the interfering excitation harmonics. Additional studies are also needed to cope with the irregular harmonic pattern of high-pitched singing voice.

%\clearpage
\bibliographystyle{IEEEbib}
\footnotesize
\bibliography{refs}

%\bibliography{biblio_short} 

\end{document}